\documentclass[draft,jgrga]{agutex}

\usepackage{rotating}

\usepackage{graphicx}
\usepackage{amsmath}
\setkeys{Gin}{draft=false}
\authorrunninghead{KIRALY-PROAG ET AL.}

\titlerunninghead{VALIDATING INDUCED SEISMICITY FORECAST MODELS}

\authoraddr{Corresponding author: E. Kir\'aly-Proag,
Swiss Seismological Service, ETH Zurich, NO H51.3, Sonneggstrasse 5, 8092 Zurich, Switzerland.
(eszter.kiraly@sed.ethz.ch)}

\begin{document}

%
%

\title{Validating induced seismicity forecast models --- Induced Seismicity Test Bench}

%
%

%
%



\authors{Eszter Kir\'aly-Proag\altaffilmark{1},
J. Douglas Zechar\altaffilmark{1}, Valentin Gischig\altaffilmark{2}, Stefan Wiemer\altaffilmark{1}, Dimitrios Karvounis\altaffilmark{1}, and Joseph Doetsch\altaffilmark{2}
}

\altaffiltext{1}{Swiss Seismological Service, ETH Zurich, Zurich, Switzerland.}

\altaffiltext{2}{Swiss Competence Center for Energy Research (SCCER-SoE), ETH Zurich, Zurich, Switzerland}

\vspace{1cm}
An edited version of this paper was published by AGU. Copyright (2016) American Geophysical Union.
Kir\'aly-Proag, E., J. D. Zechar, V. Gischig, S. Wiemer, D. Karvounis,
and J. Doetsch (2016), Validating induced seismicity forecast models --- Induced Seismicity Test Bench, J. Geophys. Res. Solid Earth, 121, doi:10.1002/2016JB013236. To view the published open abstract, go to http://dx.doi.org and enter the DOI.




%
%


\keypoints{
\item We introduce a CSEP-based objective test bench for induced seismicity forecast models.
\item We introduce a 3D smoothed seismicity model for induced earthquakes.
\item We compare forecast models with different physical and statistical elements on two EGS reservoirs.
}



%
%


\begin{abstract}
Induced earthquakes often accompany fluid injection, and the seismic hazard they pose threatens various underground engineering projects. Models to monitor and control induced seismic hazard with traffic light systems should be probabilistic, forward-looking, and updated as new data arrive. In this study, we propose an Induced Seismicity Test Bench to test and rank such models; this test bench can be used for model development, model selection, and ensemble model building. We apply the test bench to data from the Basel 2006 and Soultz-sous-For\^ets 2004 geothermal stimulation projects, and we assess forecasts from two models: Shapiro and Smoothed Seismicity (SaSS) and Hydraulics and Seismics (HySei). These models incorporate a different mix of physics-based elements and stochastic representation of the induced sequences. Our results show that neither model is fully superior to the other. Generally, HySei forecasts the seismicity rate better after shut-in, but is only mediocre at forecasting the spatial distribution. On the other hand, SaSS forecasts the spatial distribution better and gives better seismicity rate estimates before shut-in.
The shut-in phase is a difficult moment for both models in both reservoirs: the models tend to underpredict the seismicity rate around, and shortly after, shut-in. 
\end{abstract}

%
%

%

\begin{article}

%
%

\section{Introduction}
\subsection{Induced seismic hazard}
Seismicity caused by human activity, what is currently being called induced seismicity, is not a new phenomenon. Over the last several decades, workers have noted that earthquakes are triggered by human activities including nuclear explosions \citep{Boucher1969}, fluid extraction \citep{Segall1989}, fluid injection \citep{Seeber2004, Ellsworth2013}, controlled filling of artificial reservoirs (e.g., Koyna, India) \citep{Gupta2002a}, and mining and excavation \citep{McGarr1976}.
But interest in induced seismicity has recently spiked, as has the rate of induced earthquakes in the central and eastern US \citep{Ellsworth2013,Weingarten2015}. Here, it appears that fluid injections, primarily involving wastewater, are causing extensive seismic activity including events such as the 2011 $m_{w}~4.0$ earthquake in Youngstown, Ohio, \citep{Kim2013}, the 2011 $m_{w}~4.7$ central Arkansas earthquake \citep{Horton2012}, the 2011 $m_{w}~5.7$ central Oklahoma earthquake \citep{Keranen2013}, and the 2012 $m_{w}~4.9$ east Texas earthquake \citep{Frohlich2014}.

For modern deep geothermal energy projects, induced seismicity is a concern because fluids must be injected to stimulate and enhance reservoir permeability, allowing the heat to be extracted. There are two recent examples in Switzerland: the Basel EGS experiment in 2006 \citep{Haring2008} and the St.~Gallen hydrothermal injection in 2013 \citep{Kraft2013,Edwards2015,Obermann2015}. Both projects were canceled: Basel because of widely-felt seismic activity, and St.~Gallen due to gas inflow, the low natural fluid flow rate, and the high level of seismic activity during a short-term stimulation. These experiments demonstrated that project managers and operators have to be able to manage induced seismic hazard and must strike a balance between reservoir creation (i.e., permeability enhancement, which is required for a geothermal system to be profitable) and induced seismicity. Induced seismicity during geothermal projects is a blessing and a curse: the spatial extent of micro-seismicity is a proxy for the size of the stimulated reservoir, but felt and potentially-damaging earthquakes pose seismic risk to people and infrastructure. Induced earthquakes in deep geothermal reservoirs are usually smaller than $m~3$, but larger events ($>m~4$) can occur, the largest so far being an $m~4.6$ earthquake at the Geysers geothermal site in 1982 \citep{Majer2007}. Certainly, induced earthquakes felt by the public may deter future geothermal projects. 
Despite the cancellations at Basel and St.~Gallen, several geothermal projects in Switzerland are in development. As part of the Swiss national energy strategy, deep geothermal heat should supply $5-10\%$ of the national baseload electricity \citep{Giardini2014}. One of the main obstacles to achieving this goal is induced seismic hazard. To minimize induced seismic hazard, it is crucial not only to monitor and analyze induced events, but also to develop a near-real-time tool for making operational decisions. Such a hazard management scheme should be used to plan and operate reservoir stimulation so that large induced earthquakes are avoided \citep[e.g.,][]{Bachmann2011,Mena2013,Goertz-Allmann2013}. 

\subsection{Near-real-time forecasting: towards an adaptive traffic light system}
\citet{Bommer2006} introduced a traffic light system to monitor and react to seismic activity during geothermal reservoir stimulation. Like most traffic lights, this system distinguished three hazard levels, which were based on the size of events, observed peak ground velocity, and public response. But the thresholds used to change the light were chosen subjectively, primarily by expert judgment \citep{Hirschberg2015}, and in practice the system has resulted in operators taking action too late to avoid large events or a high seismicity rate. For example, in Basel the early induced earthquakes suggested that felt events were likely, but the traffic light system failed to anticipate them \citep{Haring2008}. 
An improved hazard management scheme should be a dynamic, forward-looking system that incorporates real-time data and makes probabilistic forecasts of induced seismicity and its consequences. Such an Adaptive Traffic Light (ATL) system is composed of several modules (Figure \ref{fig0}): 
\begin{enumerate}
\item Collecting prior information, e.g., geological setting for hazard assessment and building classifications for risk assessment (yellow in Figure \ref{fig0}). These data are essential to plan a geothermal project and can address questions such as where to drill wells, the orientation of the local stress field, how to design reservoir creation plans, and the maximum possible magnitude \citep{Gischig2015}.
\item Real-time data flow of hydraulic and seismic information (red in Figure \ref{fig0}). These are hydraulic data (e.g., injection flow rate and pressure measurements in the well) and seismic data that allow one to monitor reservoir creation, circulation, or other activities in the reservoir.
\item Modeling and forecasting seismicity (orange in Figure \ref{fig0}). The key element in an ATL system is seismicity forecasting. To forecast, we consider two periods: a learning period and a forecast period. During the learning period, seismic events are observed and analyzed according to their distribution in time and space. Then a calibrated model forecasts the number, magnitude distribution, and spatial distribution of events in the forecast period. 
\item Ground motion models (gray in Figure \ref{fig0}). These models estimate the shaking that an earthquake will cause and are based on properties of the earthquake source (e.g., its magnitude, style of faulting, and depth), wave propagation (distance to the earthquake), and site response (type of rock, soil that can attenuate or amplify ground shaking). Ground Motion Prediction Equations \citep{Douglas2013} and the Virtual Earthquake Approach \citep{Denolle2013,Denolle2014} are examples of possible choices to estimate ground motions.
\item Combining models to account for epistemic uncertainties (green in Figure \ref{fig0}). No single model captures all of the important features of seismicity. Model combination using appropriate weights is one way to try to leverage each model's best features.
\item Calculating hazard and risk (brown in Figure \ref{fig0}). One can estimate the seismic hazard --- the probability that some level of shaking will be exceeded --- by combining ground motion models and either synthetic catalogs generated by forecast models or individual scenario earthquakes. One can use this hazard to estimate the seismic risk: the potential economic, social, and environmental consequences of seismicity.
\item Guiding on-site decision-making processes (white  in Figure \ref{fig0}). Based on hazard and risk calculations, operators can make decisions concerning future stimulation strategies and adjust flow rate accordingly. 
\end{enumerate}

In this paper, we focus on the forecast models and the performance assessment modules of the ATL system (delineated by a dashed gray line in Figure \ref{fig0}).

\subsection{Models to forecast seismicity}
Induced seismicity models can be grouped into three classes \citep[e.g.,][]{Gischig2013a,Gaucher2015}: statistical, physics-based, and hybrid. In general, statistical models for induced seismicity \citep[e.g.,][]{Reasenberg1989,Hainzl2005,Bachmann2011,Mena2013} are conceptually and computationally simple and include aleatory uncertainty. But they do not explicitly account for the physical processes governing induced seismicity (e.g., fluid flow in fractures, permeability changes, and stress interaction) and, until this study, they have not been used to forecast the spatial distribution of earthquakes. It is sometimes thought that statistical models, because they are primarily based on clustering, are limited in their ability to predict large events or make accurate long-term forecasts. In contrast, physics-based models \citep[e.g.,][]{Olivella1994, Bruel2005, Kohl2007, Baisch2010, Rinaldi2015, McClure2012, Wang2012, Karvounis2015, Mignan2015b} do consider underlying physical processes, and are hoped to perform better when operational conditions change, such as for the shut-in period, and for long-term forecasts. But the high computational expense of most physics-based models precludes their use in near-real-time applications for the moment. 
Hybrid models are a compromise between physical models and statistical models. The goal of hybrid model development is to include some physical complexity and replace more complex physical considerations with statistical methods or stochastic processes. \\
\citet{Mena2013} compared forecast models using the Basel dataset and found that Shapiro's model \citep{Shapiro2010} provided a good fit to the rate of induced earthquakes. This model uses the seismogenic index, $\Sigma$, a parameter that describes the expected seismic response of a given site. The seismogenic index is a function of the total injected fluid volume and can be estimated from a short injection period or from the entire stimulation period; it also takes into account the $b$-value of the observed seismicity and the total injected volume. Using $\Sigma$, one can forecast the number of earthquakes in a given magnitude range and given period. Like most statistical models for induced seismicity \citep[e.g.,][]{Bachmann2011}, Shapiro's model does not make any predictive statements about the size or shape of the seismicity cloud. But it is crucial to monitor and anticipate the shape and size of the seismic cloud during reservoir stimulation for two reasons. First, the extent of the seismicity cloud is used to estimate the volume of the stimulated reservoir, which is crucial for energy production. Second, the spatial distribution of seismicity affects hazard and risk analysis: many geothermal sites are located near settlements, making energy transportation cheap but posing a risk to infrastructure and people \citep{Edwards2015}. 
Seismic risk strongly depends on geological settings (e.g., rock type under the settlement), building vulnerability, and the depth of induced events. For instance, if a $m_{w}~4$ event occurs $5~km$ below strong, new homes built on a rock site, almost all buildings would remain intact, with only some slight damage. If an event of the same size occurs $3~km$ below vulnerable houses built on a sedimentary basin, it is more likely that the houses would be slightly damaged, and some houses may be moderately or even heavily damaged \citep{Grunthal1998a}. Because the spatial distribution of induced seismicity is so important, any ATL system should be driven by 3D spatial forecasts. 

In this study, first we extend Shapiro's model to produce 3D forecasts (SaSS model, i.e., Shapiro and Smoothed Seismicity model). Then, we perform systematic statistical tests on this model and on a hybrid model, in which seismicity is triggered by a numerically modeled pressure diffusion (HySei model, i.e., Hydraulics and Seismicity model). To date these are the only models in our institute, that are calibrated against real data, and systematic re-calibration and testing can be carried out; moreover, they have a good variety of model features, which forecasts are worth evaluating and comparing. To do this, we develop an Induced Seismicity Test Bench.

\subsection{Induced Seismicity Test Bench}
Little work has been done on model selection and model comparison in the context of induced seismicity. To validate, compare, and rank models that can be used for ATL systems, we propose a model development test bench that follows the Collaboratory for the Study of Earthquake Predictability (CSEP, http://www.cseptesting.org/) approach for tectonic earthquakes. CSEP supports scientific earthquake prediction experiments in natural laboratories in multiple regions and spanning the globe \citep[e.g.,][]{Gerstenberger2010, Schorlemmer2010, Zechar2010c, Nanjo2011, Eberhard2012, Mignan2013, Taroni2013, Zechar2013}. This support comes in the form of testing centers that CSEP operates; these centers allow modelers to check the consistency of their model with observations and to compare models. We describe these activities in more detail in Subsection 3.2.

The proposed Induced Seismicity Test Bench requires models to be tested, good quality induced seismicity datasets, and a robust statistical testing framework allowing objective model evaluation. To test model consistency with observations and to rank models, we rely on pseudo-prospective forecasting, i.e., data that come from past stimulation experiments. Modelers calibrate their models using data recorded during a learning period and make forecasts for a subsequent forecast period. Since observed data of the forecast periods are already available, we can compare observed and forecast data after each recalibration and test the consistency of the forecast in terms of seismicity rate, spatial distribution, and magnitude distribution. We can use statistical metrics such as the information gain per earthquake to compare model pairs and rank models according to their forecast skill \citep{Rhoades2011}. Modelers should use the results of testing for further development, creating a feedback between testing and modeling. The long-term goal is to develop an operational ATL system to plan and conduct reservoir creation without a high rate of seismicity or large events. A detailed flowchart of the Induced Seismicity Test Bench can be found in the supplement (Figure S1).\\
The Induced Seismicity Test Bench is a diagnostic tool: it can highlight which model elements, be they physical or statistical, are essential for good forecasts, and why. This can in turn improve the models and our understanding of the underlying physical phenomena. In addition to using the test bench as a diagnostic tool, it can also be utilized on the fly to judge the performance of several models since the last forecast. The results can then be used for further improvement of the individual models and/or they can be applied to weight the models for the next forecast.\\
In the next section, we briefly describe the data from two Enhanced Geothermal Systems: the Basel 2006 experiment and the Soultz-sous-For\^ets 2004 stimulation. In Section 3 we present two models, SaSS (Shapiro and Smoothed Seismicity) model and HySei (Hydraulics and Seismicity), which are calibrated on the datasets; and we also detail the testing approach. We describe the testing results in section 4, discuss our findings in section 5, and conclude in section 6. 

\section{Data}
The data we consider in this study come from the Soultz-sous-For\^ets 2004 and Basel 2006 geothermal stimulations. \\
The Basel geothermal site is located in northwestern Switzerland, at the southeastern part of the Upper Rhine Graben (Figure \ref{fig1}.a). The graben structure is an inactive extensional rift system oriented N-S \citep{Zoback1992}. Here, the crystalline basement is covered by $2.4~km$ of sedimentary rock \citep{Haring2008}. The well BASEL1 was drilled to a depth of $5~km$ between May and October 2006. In December 2006, after several hydraulic tests, the reservoir was hydraulically stimulated to enhance its permeability. The plan was to stimulate for 21 days, but after 6 days the injection was stopped due to intensive seismicity. In the year that followed, 3 additional events of $m_{L}~>~3.0$ followed \citep{Haring2008}. Based on the results of a subsequent risk study \citep{Baisch2009,Secanell2009}, the project was abandoned. After several years, the reservoir still has earthquakes, but the seismicity rate is very low (1-3 earthquakes recorded per year) \citep{Deichmann2014}. In this study, we use about 15 days of hydraulic \citep{Haring2008} and seismic data \citep{Dyer2010} from the beginning of the stimulation (2006-12-02, 18:00), and we also use the pre-stimulation injection test data. \\
The Soultz-sous-For\^ets geothermal site is also located in the Upper Rhine Graben, between Kutzenhausen and Soultz-sous-For\^ets, about 70 km north of Strasbourg (Alsace, France; inset in Figure \ref{fig1}). The geothermal gradient is about $100^{\circ}C/km$ within the $1.5~km$ thick sedimentary cover over a granitic basement \citep{Evans2012}. This abnormally high geothermal gradient is related to deep hydrothermal convection cells in the fractured basement \citep{Gerard2006}. The geothermal project here started in the early 1980s and four wells have been drilled into two reservoirs: one at about $3.5~km$ depth (GPK1, GPK2 wells) and another at about $4.5~km$ (GPK2, GPK3, GPK4 wells). Several stimulations and circulation tests were carried out \citep{Gerard2006, Calo2013, Genter2012}. Energy production started in 2008 \citep{Genter2010}. In this study, we use hydraulic and seismic data of the pre-stimulation and stimulation of September 2004 (Figure \ref{fig1}.b, \citet{Dyer2005}). Local magnitudes were corrected by using the scaling relationship by \citet{Douglas2013}. Note that the seismograms in this data set are clipped, causing saturation of the magnitudes at $1.8$; that is, no event has $m_{w} > 1.8$.

\section{Models and testing}
\subsection{The Shapiro and Smoothed Seismicity (SaSS) model}
The SaSS model is computationally simple and based on the seismogenic index, $\Sigma$ \citep{Shapiro2010}; we distribute the earthquakes expected by $\Sigma$ in 3D by smoothing seismicity in space. Shapiro's model, which describes the rate of induced seismicity during stimulation, is defined as:

\begin{equation}
 log_{10}(N_{m}(t)) = log_{10}(Q_{c}(t))-bm-\Sigma
 \label{eq1}
\end{equation}

where $N_{m}(t)$ indicates the number of induced events above magnitude $m$ up until time $t$, $Q_{c}(t)$ denotes the cumulative injected volume of fluid at time $t$, $b$ is Gutenberg-Richter $b$-value of the observed seismicity, and $m$ is the magnitude above which all events are expected to be reliably recorded (often called the magnitude of completeness).
\\
To forecast the number of events in the forecast period, we estimate $\Sigma$ and $b$ from the learning period, and we predict the total volume that will be injected by the end of the forecast period. \citet{Kiraly2014} compared four deep geothermal datasets and found that in some cases $b$ and $\Sigma$ are not constant during and after stimulation; thus, we re-estimate them at the end of each learning period, every six hours. To predict $Q_{c}(t)$ at the end of a forecast period, we assume that the injection flow during the forecast period will follow the previously-planned strategy. Eq. \ref{eq1} describes the rate of induced seismicity only during stimulation \citep{Shapiro2010}.
As soon as the stimulation stops (the moment of well shut-in), the rate of induced earthquakes is expected to decay; the SaSS model assumes the decay follows the equation of \citet{Langenbruch2010} (using the original notation for consistency):

\begin{equation}
 R_{0b}\bigg(\frac{t}{t_{0}}\bigg) = \frac{R_{0a}}{\bigg(\frac{t}{t_{0}}\bigg)^p}
 \label{eq2}
\end{equation}

where $R_{0b}$ is the post-stimulation seismicity rate at time $t$ (since the beginning of the stimulation), $t_{0}$ is the length of the stimulation period before shut-in, $R_{0a}$ denotes the average seismicity rate during stimulation, and $p$ controls how quickly the rate decays. For subsequent forecast time windows (i.e., 6-hour time bins of the forecast period, FTWs), the majority of parameters are calibrated on the corresponding learning period, but $Q_{c}$ and $R_{b0}$ are recalculated for each time window. If the learning period ends in the stimulation period but some FTWs expand to the post-stimulation, the estimation of parameter $p$ is not possible, thus we use a generic value: $p = 2$. Also, if $p$ is estimated to be smaller than 2 we set the value to 2, following the value that is proposed by \citet{Langenbruch2010} for an early post-injection period. Detailed flowchart of number component can be found in the supplement (Figure S2).
As in CSEP experiments and suggested by \citet{Shapiro2010}, the number of events in each forecast period is assumed to follow a Poisson distribution and the numbers obtained by using Eq. \ref{eq1} and \ref{eq2} are Poisson expected values; error bars in all subsequent figures indicate the $95\%$ Poisson confidence interval.
\\
To model the 3D spatial distribution of induced earthquakes, we added a spatial component to the model by smoothing the seismicity observed during the learning period (Figure \ref{fig2}.A). Several studies, including the Regional Earthquake Likelihood Models (RELM) experiment \citep{Schorlemmer2010,Zechar2013} have shown that smoothed seismicity models are effective at forecasting the spatial distribution of tectonic earthquakes. To construct a smoothed seismicity model in two dimensions, one applies a two-dimensional smoothing kernel to each past event \citep[e.g.,][]{Helmstetter2007}, calculates the contribution of smoothed earthquakes on a given grid, then sums contributions of all observed earthquakes. To create a probability density function (PDF, i.e., earthquake spatial probability map), one normalizes the smoothed seismicity map so its sum is unity.

We extend the 2D Gaussian smoothed seismicity model of \citet{Zechar2010b} to 3D. For each forecast period, we smooth all prior events, where the contribution of an earthquake to a given voxel (i.e., volume element) is

\begin{eqnarray}
 K(x_{e},y_{e},z_{e},x_{1},x_{2},y_{1},y_{2},z_{1},z_{2}) & = &
 \frac{1}{8} \Bigg[ erf \bigg( \frac{x_{2}-x_{e}}{\sigma_{1} \sqrt{2}} \bigg) - erf \bigg( \frac{x_{1}-x_{e}}{\sigma_{1} \sqrt{2}} \bigg) \Bigg] \nonumber \\ 
&& \times \Bigg[ erf \bigg( \frac{y_{2}-y_{e}}{\sigma_{2} \sqrt{2}} \bigg) - erf \bigg( \frac{y_{1}-y_{e}}{\sigma_{2} \sqrt{2}} \bigg) \Bigg] \nonumber \\
&& \times \Bigg[ erf \bigg( \frac{z_{2}-z_{e}}{\sigma_{3} \sqrt{2}} \bigg) - erf \bigg( \frac{z_{1}-z_{e}}{\sigma_{3} \sqrt{2}} \bigg) \Bigg]
\label{eq3}
\end{eqnarray}

where $x_{e}$, $y_{e}$ and $z_{e}$ denote the location of the given earthquake, $x_{1}$, $x_{2}$, $y_{1}$, $y_{2}$, $z_{1}$ and $z_{2}$ are the points that define the edges of the voxel, and $\sigma_{1}$, $\sigma_{2}$ and $\sigma_{3}$ are bandwidths of the 3D Gaussian kernels in EW, NS and vertical directions, respectively.
To make a good smoothed seismicity forecast, we need good bandwidths; we optimize these by dividing data from the current learning period into a training set and a validation set (Figure \ref{fig2}.C). The length of the training and validation sets depend on the length of the forecast period and the learning period. If the length of the forecast period is more than half the length of the learning period, the training and validation sets are each one-half of the learning period. Otherwise, the length of the validation set is equal to the length of the forecast period. We search for the bandwidth combination that, when used to smooth the training set, best forecasts the seismicity of the validation set. To avoid 'surprises,' i.e., events occurring where the model would not expect any events, we distribute a certain fraction of the PDF over all voxels (i.e., surprise factor), following the idea of \citet{Kagan2000}.
We analyze the performance of $1000$ combinations of bandwidths and surprise factors using the training and validation set of the learning period.
The PDF is updated for each new learning/forecast period. Since the PDF is based on the learning period, this model assumes that earthquake locations in the forecast period will not be very different from the seismicity observed so far.

Smoothed induced seismicity models must differ from their tectonic counterparts in at least one aspect: induced models should capture the propagation of the seismicity front after shut-in. In particular, due to pore pressure diffusion, induced seismic activity tends to decrease in the vicinity of the injection well and to concentrate at the boundaries of the reservoir. We attempt to model this time-dependent effect by applying exponential temporal weighting: the most recent event receives a maximum weight (one), and earlier events get smaller weights according to their origin time. This is analogous to the exponential smoothing approach commonly used in time series forecasting \citep{Goodwin2010} and is also connected to the Omori-Utsu relation describing aftershock decay rate \citep{Zhuang2012}.

The forecast magnitude distribution is the Gutenberg-Richter distribution \citep{Gutenberg1944} with the $b$-value estimated from the learning period.

\subsection{The Hydraulics and Seismicity (HySei) model}

The HySei model developed by \citet{Gischig2013a} describes seismicity triggered by pressure diffusion with irreversible permeability enhancement. The biggest advantage of the model is that it quantifies permeability enhancement by calibrating flow rate and wellhead pressure against observations. 
The HySei model consists of two main parts: hydraulic inversion and seismicity modeling.
The aim of inverting hydraulic observations is to reconstruct the pressure evolution in the reservoir. We seek the best hydraulic parameters to match the observed well-head pressure with a one-dimensional radial flow model. We use a finite difference method in a circle of $1200~m$ radius distributed on $3000$ nodes, and $1$-minute resolution in time. During the pre-stimulation test injection, we solve the diffusion equation (Eq. \ref{eq1a}) with constant permeability ($\kappa = \kappa_{0}$). During stimulation the governing equations are the diffusion equation (Eq. \ref{eq1a}) with irreversible changing permeability (Eq. \ref{eq1b}) due to increasing pressure that exceeds some threshold (Eq. \ref{eq1c}):

\begin{equation}
\rho S \frac{\partial p}{\partial t} = \nabla \Big( \frac{\kappa\rho}{\mu} \nabla p \Big) + q_{m}
\label{eq1a}
\end{equation}

\begin{equation}
\kappa = \kappa_{0} (u + 1)
\label{eq1b}
\end{equation}

\begin{equation}
\frac{\partial u}{\partial t} = C_{u} H_{pt}\Big( \frac{\partial p}{\partial t} \Big) H_{u} (u_{t}-u)H_{p}(p-p_{t})
\label{eq1c}
\end{equation}

where $\rho$ is fluid density, $S$ is the specific storage coefficient, $\kappa$ is permeability that varies during the stimulation, $\mu$ is fluid viscosity, and $q_{m}$ is a mass source; $\kappa_{0}$ is the initial permeability before the stimulation, $u$ is stimulation factor (i.e., the overall permeability enhancement of the reservoir); $C_{u}$ is stimulation velocity, a constant that scales the rate at which permeability changes, $u_{t}$ is maximum stimulation factor, and $p_{t}$ is threshold pressure, $H_{pt}$ is a Heaviside function, it is one if pressure increases, zero otherwise, $H_{p}$ and $H_{u}$ are Heaviside functions for pressure and stimulation factor. These are smoothed to avoid a singularity and resulting numerical instability. Permeability starts to increase if pressure reaches $p_{t}$. If pressure further increases, the permeability of the reservoir increases until it reaches $u_{t}$. Note that a reversible component of permeability change representing the compliant response fracture to pressurization \citep[e.g.,][]{Rutqvist2003} has not been included in this version of the model.

In the seismicity model, randomly-placed potential nucleation points are triggered by the radial symmetric pressure evolution following the Mohr-Coulomb failure criterion. They have no spatial extent, but differential stress ($\sigma_{1}-\sigma_{3}$) is defined at the seed point. Local $b$-values are determined at the seed points following a linear relationship between differential stress and $b$-value: $b_{max}$ and $b_{min}$ parameters are $b$-values at minimum and maximum values of differential stress, respectively. When a seed point is triggered, a random magnitude is drawn from the magnitude distribution with the local $b$-value. Additional free parameters are the scaling factor $F_{s}$ (the ratio between the number of synthetic and observed events), the stress drop coefficient $d\tau$ (the change of stress conditions after a seed has been triggered), and a criticality threshold $d\mu$, which accounts for the fact that seed points cannot be too close to the failure limit. \\
For this study, we parallelized parts of the code and extended the model to 3D (Figure \ref{fig2}.B) by adding an off-fault component to the originally 2D seismicity model. Assuming that the seismicity is generated on the current main fault, we determine the principal components of the current seismicity cloud and use the empirical distribution of the seismicity along the smallest axis to define off-fault coordinates of the synthetic events.
A detailed flowchart of the HySei model can be found in the supplement (Figure S3).
\\
To represent the spatial differences of the two models, Figure \ref{fig3} shows cross sections of the 3D PDFs of SaSS (upper line) and HySei (bottom line) at the moment and location of the biggest event ($m_w~=~3.1$), which occurred about $5$ hours after the shut-in.  

\subsection{Testing}
To assess a single model, we check if its forecasts are consistent with the observations \citep{Zechar2010a}, asking the question: might the observations have been generated by this model? One way we do this is to check if the number of observed earthquakes falls within the $95\%$ confidence interval of the forecast. If so, the model passed the Number-test. In a similar way, we examine if the magnitude distribution of all forecasts is consistent with the observations (Magnitude-test). To test the spatial component (Space-test) \citep{Zechar2010a, Rhoades2011}, we use a testing grid of $4 km \times 4 km \times 4 km$ centered on the well tip and divided into $200 m \times 200 m \times 200 m$ voxels. After normalizing the forecasts so that the number of forecast events matches the number of observed events, we calculate the log-likelihood (LL) of the observation in each voxel. Summing these values gives a joint LL for a specific experiment. The higher the joint LL values are the better the forecast \citep{Zechar2010a, Rhoades2011}.

To check if the forecast is consistent with the observed seismicity of the forecast period, we simulate $1000$ catalogs from the forecast, and find the $5^{th}$ percentile of the LL values for the simulated catalogs. If the LL for the current observation is higher than the $5^{th}$ percentile the forecast passed the Space-test --- the observed seismicity could have been generated by the model. Both models consider the earthquake distribution Poissonian, thus LL values are calculated as follows:

\begin{equation}
 L(A) = \sum\limits_{i=1}^{n} \Big[k_{i} \times log \big(\lambda_{A_{i}} \big)-\lambda_{A_{i}}-log \big(  k_{i}! \big) \Big]
\label{eq4}
\end{equation}

where $L(A)$ is the Poisson joint LL of forecast A, $n$ is the number of voxels, $k_{i}$ is the number of earthquakes observed in the ith voxel, and $\lambda_{A_{i}}$ is the forecast seismicity rate in the $i^{th}$ voxel of forecast A. 
\\
To compare two models, one can directly compare individual LL values of the models either for model components (i.e. event numbers, magnitudes or the spatial component) separately or for the entire model. These measures give information about the model performance not only against data but against other models. Here we would like to emphasize that LL values consider the whole model space. In other words, it reflects the performance of not only the temporal/magnitude/spatial bins that host at least one earthquake but also the empty ones answering the question: what is the probability to have zero earthquake in the given temporal/magnitude/spatial bin? 

One can also calculate the information gain of one model with respect to another for model comparisons. This measure emphasizes the non-empty bins by comparing the forecast seismicity rates of model $A$ with that of model $B$ in the voxels where earthquakes occurred. The following formula gives $I_{i}$, the information gain of model $A$ over model $B$ for an earthquake occurring in the $i^{th}$ voxel \citep{Rhoades2011}:

\begin{equation}
 I_{i} = \frac{-N_{A}+	N_{B}}{N} + ln \Bigg( \frac{\lambda_{A_{i}}}{\lambda_{B_{i}}} \Bigg)
 \label{eq5}
\end{equation}

where $N$ is number of observed events, $\lambda_{A_{i}}$ and $\lambda_{B_{i}}$ denote forecast seismicity rate in the $i^{th}$ voxel of model $A$ and $B$, respectively, $N_{A}$ and $N_{B}$  are the total forecast number of events in model $A$ and $B$, respectively. The first term of the right hand side is a penalty concerning the number of events under each model. We seek to know if one model is better than the other, in other words, if the expected value of the information gain population differs from zero. One can also estimate how much better or worse model $A$ relative to model $B$ (i.e., average information gain) by finding an appropriate estimator. Exponentiating the average information gain yields the average probability gain of model $A$ with respect to model $B$. Additionally, $95\%$ confidence interval of the estimated expected value can be calculated to determine if model $A$ is significantly better or worse than model $B$: if the confidence interval contains zero, the difference between the models is not statistically significant at the $5\%$ significance level. \\
Several techniques are possible to compute the average information gain. \citet{Rhoades2011} suggested to take the arithmetic mean of the information gain distribution as the expected value of the population, based on Student’s t-distribution \citep{Student1908}. We refer to this method as 'Classical mean'. This estimator is best if the population follows a normal distribution. Plotting the distribution of information gains (that is, for individual earthquakes) for SaSS relative to HySei as a function of time and in a quantile-quantile plot (Figure S4) suggests that the information gains are not normally distributed. One possible way to solve this problem is to seek an estimator that can tackle outliers systematically. This can be done by manual data screening and removal of outliers, but it can be impractical due to the large number of data points and possible masking (i.e., large outliers can hide smaller ones). To overcome these problems, we use robust statistics to automatically detect and downweight outliers \citep{Ruckstuhl2014}. We refer to this method as 'Robust mean'. To calculate the expected value of the information gain distribution, we compute a weighted mean where the influence of the outliers is reduced. In particular, we use the Huber M-estimator, implemented as \textit{mlochuber} in the LIBRA matlab package \citep{Verboven2005}. By using the Huber M-estimator, we avoid the problem that a few earthquakes dominate the estimate of the average information gain.
We also explore a non-parametric method: generate $1000$ bootstrap samples of the observed information gains (i.e., we sample with replacement) and find the arithmetic average and $2.5\%$ and $97.5\%$ percentiles, thus obtaining a "Bootstrap mean" and the corresponding $95\%$ confidence interval. Using the same bootstrap samples we also find 'Bootstrap median'. We show a comparison of these methods in the next section.

\section{Results}
\subsection{Consistency tests}
Figure \ref{fig4} shows four snapshots of forecast and observed seismicity rates for both datasets. The top row shows the corresponding hydraulic data (injection rate and well-head pressure) to provide time reference for the forecasts. Blue, red, green, and purple vertical lines indicate the end of the different learning periods: corresponding shaded areas show forecasts of SaSS model (middle row) and HySei model (bottom row) with $95\%$ Poissonian confidence intervals. In case of Basel 2006, both models seriously overpredicts the seismicity rate for LP1 (blue learning period that ends at day $1.25$). This might be due to the short learning period. Giving longer learning period to the models (LP2, red learning period that ends at day $3.25$), the forecast is greatly improved for both models. SaSS struggles to forecast after both LP3 (green learning period that ends at day $5.25$) and LP4 (purple learning period that ends at day $9.5$), while HySei underpredicts after LP3 and gives perfect forecast after LP4.
In the case of Soultz-sous-For\^ets 2004, SaSS gives good forecasts at first (after LP1, the learning period that ends at day $1.75$), then severely underpredicts (after LP2 the learning period that ends at day $3.5$), and finally significantly overpredicts the seismicity rate (after LP3 and LP4, the learning periods that end at day $5$ and $6.5$, respectively). HySei performs well in most of the cases (after LP2, LP3 and LP4), except after LP1. In this case, the model expects higher pressure in response to the injection peaks between day $2-3$, which results in overprediction of the sesimicity rate. This might be due to the fact that a reversible component of permeability change, possibly arising from fracture compliance, is not included in this version of the model.

To show forecasts corresponding to all learning periods, we use a matrix representation where colors indicate the goodness of the forecast (Figure \ref{fig5}): yellow means a perfect forecast; red and blue mean under- or overprediction, respectively. Downward- and upward-pointing triangles denote moments when the observed seismicity rate falls out of the $95\%$ confidence intervals due to serious under- or overprediction, respectively. To avoid overlap of the forecast periods, we represent the $3$-day forecast period vertically: the end of the learning period is indicated on the horizontal axis, time during the $3$-day forecast period is indicated on the vertical axis with subsequent $6$-hour FTWs. Time in the forecast period increases from bottom to top. 
The top row of Figure \ref{fig5} shows the observed seismicity rate for both datasets, middle and bottom rows show a comparison of observed seismicity rates with forecasts from SaSS and HySei, respectively. 
In Basel, both models mainly overestimate the number of observed earthquakes during the initial stimulation period. When the injection rate was decreased and at shut-in, both models have difficulties forecasting the right number of earthquakes: they severely underpredict the observed seismicity rate. The SaSS model overpredicts for the post-stimulation period, whereas HySei seems to find good estimates most of the time for later periods (with the exception of three time windows). In Soultz-sous-For\^ets 2004, the SaSS model mainly forecasts well or overestimates the number of earthquakes during stimulation. The forecast period corresponding to the learning period of day $3.5$ stands out, when SaSS significantly underpredicted the number of earthquakes. This is because there is not yet enough data of the post-injection period to estimate post-stimulation parameters. During the post-stimulation period, the SaSS model overpredicts almost all FTWs. On the other hand, the HySei model gives generally good results: there are only a few under- and overpredictions, mainly at the beginning of the injection, around shut-in, and near the end of the investigated period. 
Overall, in most of the cases, HySei is better at forecasting the number of induced earthquakes; this is reflected by the number of unmarked FTWs in Figure \ref{fig5}. Moreover, for a small period of re-injection in Soultz-sous-For\^ets (at day $8$), HySei forecasts the number of events well, while the SaSS model significantly overpredicts.

In Figure \ref{fig6} we compare the observed magnitude distribution with forecasts from SaSS and HySei. Magnitude bins are $0.1$ units wide and range from $0.9$ to $4$ for Basel 2006 and from $0$ to $1.9$ for Soultz-sous-For\^ets 2004. We remind the reader that the Soultz-sous-For\^ets 2004 magnitudes are truncated, so the final magnitude bin contains all events that would have $m > 1.8$.
Both models forecast the magnitude distribution of micro-seismic events well, meaning that observed seismicity follows the Gutenberg-Richter relation in almost all cases. Nevertheless, the probability of the biggest event of the Basel 2006 project is very small in both models (insets in Figure \ref{fig6}b-c). The truncated magnitudes in Soultz-sous-For\^ets 2004 preclude us from considering the probability of the largest event in this data set, because we have no good estimate for the magnitude of the largest event.

We investigate the spatial component of the models by dividing the joint LL by the number of observed events (LL/Eqk) in Figure \ref{fig65}. We decided to normalize due to the fact that LL values are correlated with the number of earthquakes in a FTW. We use the same matrix representation as we introduced for the number component: end of learning periods are indicated on the horizontal axis, FTWs on the vertical axis. Yellow indicates better results than red, the higher the LL value, the better the forecast is. Crosses represent moments when the model does not pass the Space-test. Gray squares denote moments when no earthquake occurred. Gray dotted line marks the shut-in moment. It is clear that SaSS passes the Space-test more often than HySei does, especially after shut-in for both datasets. Additionally, SaSS's LL values are higher than that of HySei indicating that smoothed seismicity outperforms the simple geometry of HySei's forecasts.

\subsection{Ranking}
To be able to compare the two models we calculate LL from the absolute values of the Number- and Magnitude-test by answering the same question we addressed in case of the spatial component: what is the probability of the observation given the model forecast? We calculate LL values for all FTWs of all model components (Figure S5-S6). Figure \ref{fig7} gives an overview of differences between the model LLs. Green shows when SaSS performs better than HySei, pink shows when HySei is better than SaSS, white indicates that the models forecast similarly.
The magnitude component is exceptional in this figure, because we do not test the consistency of the forecast and observations in incremental FTWs, rather the cumulative distribution. For instance, in case a $3$-day magnitude test we take all events occurred in the forecast period from the end of the learning period until the end of day $3$. This allows a more stable distribution of the observed events that can be tested against a power law.
 
These results clearly confirm that the magnitude component is very similar in the models, which is not surprising since both models use the Gutenberg-Richter relation. The differences lay in the number and spatial components. In terms of number, SaSS performs better in several moments during the stimulation and in the early post-stimulation period in Basel. HySei gives better results close to the shut-in and generally after the stimulation, especially at later moments of the experiment. The green color in most FTWs of the spatial component reveals that SaSS holds the better spatial component, which is emphasised towards the end of the experiment.

To compare the entire model performance, we merge all components and calculate LL normalized by the number of earthquakes occurred in the given FTW. Figure \ref{fig8} details the sum of LL/Eqk values of the individual FTWs for $6$-, $24$-, $48$-, and $72$-hour forecast periods. Three regimes can be observed in the case of Basel 2006:
\begin{itemize}
\item regime $A$: when models perform similarly well
\item regime $B$: when SaSS model is better than HySei
\item regime $C$: when HySei overcomes SaSS, especially for the longer forecast periods.
\end{itemize}
Comparing these results to the performance of individual model components, it is clear that the regimes are determined by the interplay of the number and spatial components. Both components of both models perform similarly in regime $A$, which results in similar overall performance. Around the shut-in, even if HySei gives better number forecasts for a short period, SaSS can compensate with its spatial component and it overcomes HySei also with its number component by the end of regime $B$, which results in a better overall performance of SaSS for this period. As the number of events drastically decreases relative to previous periods in regime $C$, it seems that HySei's more precise number forecasts compensate against SaSS's better spatial forecasts giving better overall LL values.
In the case of Soultz-sous-For\^ets 2004, only two of the three regimes are present: regime $B$ from the beginning of the experiment about $1.5$ days after the shut-in (almost at the same moment as in Basel) and regime $C$ for the rest of the experiment. In the first part of regime $B$, the slightly better spatial component of SaSS compensates the generally better number component of HySei giving marginally better results to SaSS. From the shut-in until the end of regime $B$, the spatial component of SaSS is clearly better together with the fact that HySei's number component is less dominant than previously. This results in a drop of overall LL. The decrease of number of induced earthquakes (regime $C$) highlights again that HySei's number component overcomes SaSS's better spatial component.

Summarizing the model comparison based on LL: SaSS obtains better results in space generally, in terms of seismicity rate in some moments of the stimulation, and also the entire SaSS model gives better results until a certain point after shut-in (regime $B$) for both datasets; HySei outperforms SaSS in seismicity rate forecast in the post-stimulation period and also the overall LL values of HySei in the late post-stimulation period, especially for longer forecast periods.

Figure \ref{fig9} presents the results of all $6$-hour information gains from the beginning until the end of the experiment for both datasets. Solid black lines indicate the empirical probability densities of the information gains, dotted gray lines denote normal distributions, where the expected values and standard deviations are estimated from the corresponding empirical distributions. To use the classical method to determine the average information gain, the population should be normally distributed. This is not the case, which is why we investigate four methods to calculate the average information gain: classical mean, robust mean, bootstrap mean, and bootstrap median corresponding to red, green, orange, and brown, respectively. Insets show the estimated average values with their uncertainties. \\
For both datasets medians and robust means are closer to the the clear peaks of the populations, whereas classical and bootstrap mean values are shifted and have wider $95\%$ confidence intervals. In the case of the Basel 2006 data, interpretation of model performance depends on the choice of the estimator: for robust mean and bootstrap median HySei performs significantly better than SaSS, for classical and bootstrap mean exactly the opposite. This emphasizes that we should be cautious about information gain interpretations. 
In our opinion, in case of information gain calculations, ($1$) it is necessary to check the distribution of the observed information gains, ($2$) it is recommended to use several estimators to have a clearer view of the possible average information gain values, and ($3$) to interpret the results carefully.
An overview of average information gain for 6-, 24-, 48, and 72-hour forecast periods with all four estimators can be found in the supplement (Figure S7-S10).

\section{Discussion}

Predictive models of induced earthquakes can help reduce seismic hazard and risk during reservoir stimulations. Although many models are being developed, most are presented in a context that is descriptive, not predictive: they are tuned using the entire data set, and so their ability to forecast is not checked. In this study, we propose a test bench to objectively evaluate various induced seismicity models. We bring to the test bench two models used to forecast two datasets. We demonstrate that such a test bench can quantify the forecast skill of different models. The results can give guidance how to merge models. One possible way to combine models is weighting models by their past performance. The test bench can provide detailed information about the performance of the tested models that can be converted to probabilistic weights. Weighted average models has the potential to merge the best forecasting features of the tested models and can give important input for real-time forecasting and hazard assessment. The test bench can also highlight model features to be improved, e.g., because the model performs badly at forecasting one of the key parameters (i.e., event number, magnitude distribution, or spatial distribution) or during certain moments (e.g., during stimulation, at shut-in, or after shut-in). 

Our test bench showed that both tested models are limited to accurately forecast the rate of induced earthquakes. The forecasts are particularly bad around shut-in. During stimulation and shortly after shut-in, we observe first a slight overprediction and then a severe underprediction as the injection rate decreases and stops. In the post-injection period, SaSS overpredicts the number of events (except the moment when model parameters are not well calibrated due to the very short post-injection period). 

As suggested by \citet{Langenbruch2010}, we use a generic value of $2$ for parameter $p$ when parameter estimation is not possible, and the same generic value is used if calculated ones are lower than $2$. In Basel, we observed that calculated values of $p$ are always smaller than $2$. This means that we always apply a decay with $p = 2$, which results in faster decay than the data of learning period would suggest. Nevertheless, all modeled decays are slower than the observed seismicity decay, indicated by massive overpredictions in the post-stimulation periods. In contrast, for Soultz-sous-For\^ets estimated values of $p$ are always higher than $2$ allowing good forecasts at the beginning of the post-stimulation period but the decreasing tendency of the values of $p$ results in overpredictions for later forecast periods. These results suggest that forecasting the post-injection seismicity is difficult and the current post-injection seismicity decay law is not appropriate in an operational forecasting environment.

The spatial forecasts of the SaSS model gave generally good results. But these forecasts are limited by the fact that they are based on the current learning period. The model can give good forecasts when the seismicity is nearly stationary, i.e., new earthquakes occur where previous ones occurred. But this is often not the case in induced seismicity related to geothermal reservoir creation, where seismicity propagates with the pressure front. In future work, to incorporate diffusion-like propagation of the seismicity, we imagine a step-by-step spatial forecast for each FTW of the forecast period. One could simulate thousands of synthetic catalogs for the first FTW based on the learning period. Forecasts of FTWs are based on the PDF calculated from the synthetic catalogs of the previous FTWs. Temporal weighting (exponential or some other temporal weighting) of generated earthquakes can help to simulate the migration of the seismicity cloud.

One might also improve induced seismicity forecasting by considering Coulomb stress changes, which has been shown to a good descriptive model of tectonic seismicity \citep{Steacy2005} and has been considered in the induced seismicity context: \citet{Orlecka-Sikora2010} suggested that static stress transfer can have an accelerating impact on mining-induced seismicity, and \citet{Schoenball2012} concluded that static stress change does not play an important role during stimulation but might help to trigger after shut-in in the Soultz-sous-For\^ets reservoir. Moreover, \citet{Catalli2013} found that $75\%$ of the analyzed induced earthquakes (based on \citet{Deichmann2009}) in Basel occurred in regions of increased Coulomb stress, where failure is thought to be encouraged. Unfortunately, prospective tests of the Coulomb stress hypothesis are difficult because one needs accurate, real-time estimates of hypocenter, magnitude, and focal mechanism, and one also needs some a priori knowledge on fault orientations in the reservoir.

Additional model improvements may relate to the statistical description of earthquake distributions. In the testing framework and also in all CSEP experiments, earthquake occurrence is considered as a Poissonian process \citep{Eberhard2012}; LL and confidence interval computations are based on that assumption. The Poissonian assumption is not completely fulfilled, because earthquakes are not independent, neither in time nor in space. \citet{Eberhard2012} reported that Poissonian distribution was not supported by the seismic data; others \citep[e.g.,][]{Kagan2010,Lombardi2010} have previously shown the same observation in different regions and magnitude ranges. Failures of model forecasts might stem from the Poissonian assumption beside the fact that the model does not incorporate the necessary physical processes. Modeling earthquake occurrence as a Poissonian process is thus not ideal and improvements are subject of further investigations. 

It is necessary to emphasize that all tests are highly dependent on the observed catalog. Thus, it is extremely important to detect events and to determine good origin times, magnitudes and precise locations. For the moment, it is still a challenge, especially in near real-time. 

Our analysis further revealed that forecasting the rate and magnitude distributions around shut-in also remains a difficult question: the models often underpredict during this period and do not represent the magnitude distribution well. Presumably, this problem is not specific to the data we considered here because in several other projects the biggest event occurred after shut-in \citep{Baisch2006,Asanuma2005}. Focusing on shut-in and the events that follow, \citet{Barth2013} showed theoretically and also confirmed with the analysis of the data from Soultz-sous-For\^ets 2000 that probability of exceeding a certain magnitude can be higher after shut-in than it would have been for on-going injection. \citet{Segall2015} proposed a descriptive model that includes complete poroelastic coupling --- changes in pore pressure induce stresses, and changes in mean normal stress induce changes in pore pressure --- and concluded that an abrupt shut-in can produce sharp increase in the seismicity rate. Post shut-in peaks of the seismicity rate result from the rapid change in stress before the pore pressure can be relieved. Concerning post shut-in magnitudes, \citet{Segall2015} claimed that larger events are absent at short injection times but as injection proceeds the probability of larger earthquakes increases, thus larger events occurring post shut-in are not unexpected. Another explanation for large post-stimulation events came from \citet{McClure2015a}: simulation with the three-dimensional version of CFRAC \citep{McClure2012a} revealed that post-stimulation seismic events can be caused by backflow from dead-end fractures into fractures that host the largest event. He proposed that pumping of fluid to the surface immediately after shut-in could mitigate this effect and reduce post-stimulation seismic activity. The inferences made from these descriptive models ought to be used in future work to improve predictive models such as those considered in this study.

\section{Conclusions}
Forward-looking, near-real-time warning systems can help avoid large induced earthquakes and keep micro-seismicity at a tolerable level during and after project operations. The Induced Seismicity Test Bench can be used to test the core of such a warning system, an Adaptive Traffic Light system. Here, we tested, compared and ranked the performance of the SaSS and the HySei models.\\
To say which of these models performs best is not straightforward.  In terms of magnitude, both models forecast micro-seismicity fairly well, but none of them is able to forecast the biggest $m_{w}3.1$ event. In terms of seismicity rate, the HySei model gives good forecasts most of the time, especially for late post-stimulation periods but it can under- and overpredict at some moments. In the case of the Basel 2006 project, we observe a clear distinction between model performance: SaSS is better at some moment of the stimulation period and shortly after shut-in; HySei outperforms SaSS close to shut-in and for the most of the post-stimulation period. In terms of spatial distribution, smoothed seismicity based on learning periods (SaSS model) appears to outperform the radially symmetric geometry (HySei model). If we compare the entire models, SaSS seems to give higher LL/Eqk values at the beginning until a certain moment after shut-in when HySei takes over, especially for longer forecast periods. \\
Although our analysis is restricted to only two geothermal projects, we can generally conclude that the seismogenic index forecasts the earthquake rate better during stimulation and HySei gives better seismicity rates after shut-in; smoothed seismicity with temporal weighting performs better in forecasting the spatial component. Certainly, it would be beneficial to consider additional models and datasets in future work. In this study we introduced a comprehensive test bench for induced seismicity with the goal to better understand the behavior of injection-related reservoirs and to develop an operational Adaptive Traffic Light system for geothermal projects. With the establishment of this test bench, we challenge modelers to make predictive models, forecast induced seismicity, test their models for consistency, and compare model performance: we believe this is the most efficient way to reduce induced seismic hazard.


%
%
%
%
%
%
%

\begin{acknowledgments}
We acknowledge the GEOTHERM, GEOTHERM-$2$ and GEISERS projects for financial support to develop fundamental ideas concerning the Adaptive Traffic Light System and providing the stimulation data of Soultz-sous-For\^ets 2004. The authors would like to thank EEIG Heat Mining for permission to publish the data. Acknowledgement is also due to the numerous agencies which have supported the Soultz project over the years including the European Union, ADEME of France, BMU of Germany, SER and SFOE of Switzerland, and the EEIG 'Exploitation Mini\`ere de la Chaleur' consortium. Access to the data is provided by contacting the authors. We thank Arnaud Mignan, Antonio Pio Rinaldi and Eduard Kissling for their valuable comments on an earlier version of the manuscript. We also thank Yehuda Ben-Zion as editor, the associate editor, Carsten Dinske and three anonymous reviewers for their comments and suggestions. E.K.-P. acknowledges the GEOTHERM-$2$ project for financing her PhD. This work has been partially completed within the Swiss Competence Center on Energy Research - Supply of Electricity, with the support of the Swiss Commission for Technology and Innovation.
\end{acknowledgments}

\end{article}
%
%
%
%
%
%

%
%

%

\begin{figure}[h!]
	\caption{Flowchart of the Adaptive Traffic Light System. GMM stands for Ground Motion Models, w denotes weighting, PSHA means Probabilistic Seismic Hazard Assessment. The dotted gray line delineates the scope of this paper.}
\label{fig0}
\end{figure}

\begin{figure}[h!]
	\caption{Seismicity of the Basel 2006 (\textbf{a.}) and Soultz-sous-For\^ets 2004 (\textbf{b.}) geothermal project in NS cross sections. Wells are represented by black lines. Dotted light gray grids indicate voxels of $200m \times 200m \times200~m$ for testing. Colors denote moment magnitudes of the events, note the different scales. Map inset shows the location of the geothermal sites.}
\label{fig1}
\end{figure}

\begin{figure}[h!]
	\caption{\textbf{A.} Explanation of the spatial component of the SaSS model. \textbf{a.} General equation of smoothed seismicity. \textbf{b.} Seismicity of a learning period, colors denote temporal weights indicated in the inset. \textbf{c.} 3D Gaussian kernel represented in 2D with gray dashed and solid black lines. \textbf{d.} Vertical cross section of smoothed seismicity, colors denote the spatial probability density function. \textbf{B.} Explanation of the HySei model. Black circles represent simulated seismicity on the fault plane. Red dots indicate observed seismicity of the learning period. Minimum, maximum, and intermediate principal axes are $a_z$, $a_x$, and $a_y$ respectively. Solid black lines indicate the length of the principal axes. Black ellipse shows the $95\%$ of the seismicity cloud. Red curve shows the empirical event distribution along the minimum principal axis (i.e., off-plane direction). \textbf{C.} Explanation of time periods used for model calibration and forecasts.}
\label{fig2}
\end{figure}

\begin{figure}[h!]
	\caption{Cross sections of the SaSS (upper panels) and HySei (lower panels) forecasts for the period containing the largest event in the sequence a few hours after the shut-in (2006-12-08 16:48, $m_w3.1$). Left, middle, right panels show map view, NS vertical cross section, and EW vertical cross section at the location of the event, respectively. Black dots denote the event. Color scale indicates the probability density function of the forecast.}
\label{fig3}	
\end{figure}

\begin{figure}[h!]
	\caption{Forecast of the number of events with different learning periods. \textbf{a.} Evolution of injection flow rate in l/s (black line) and wellhead pressure in MPa (orange line) during the investigated time period of the Basel 2006 project. Dotted gray line indicates the shut-in. Blue, red, green and purple horizontal lines correspond to the learning periods starting at the begining of the injection, ending after $1.25$~days, $3.25$~days, $5.25$~days, and $9.50$~days, respectively. \textbf{b.} Number of events in $6$-hour time bins in function of time for SaSS model. Black dots represent the observed seismicity rate. Blue, red, green and purple vertical lines correspond to the previously mentioned learning periods. Shaded areas indicate the corresponding $72$-hour forecasts with $95\%$ Poissonian confidence intervals. Dotted gray line indicates the shut-in. \textbf{c.} Same as \textbf{b.} for HySei model. \textbf{d.} Same as \textbf{a.} for the Soultz-sous-For\^ets 2004 project with different moments of forecasts: blue, red, green, and purple lines correspond to the learning periods of day $1.75$, $3.5$, $5$, and $6.5$, respectively. \textbf{e.} Same as \textbf{b.} for the Soultz-sous-For\^ets 2004 project. \textbf{f.} Same as \textbf{c.} for the Soultz-sous-For\^ets 2004 project.}
	\label{fig4}
\end{figure}

\begin{figure}[h!]
	\caption{Number-test. Horizontal axes denote the length of learning periods used for forecasting, vertical axes denote individual $6$-hour forecast time windows. \textbf{a.} Observed seismicity rate of the Basel 2006 project. Color corresponds to number of observed events in $6$-hour time bins. Dotted red line indicates the shut-in moment. Blue, red, green and purple rectangles indicate forecasts (vertical direction) with learning period of $1.25$~days, $3.25$~days, $5.25$~days and $9.50$~days, respectively. These forecasts are explicitly plotted on the previous figure. \textbf{b.} Difference between number of forecast events by SaSS model and the number of observed events. Blue and red show moments when models overestimate and underestimate the observed seismicity rate, respectively. Yellow indicates similar number of forecast events as observed earthquakes. Gray upward-pointing triangles and solid downward-pointing denote moments when the number of forecast events (with Poissonian error bars) are significantly higher or lower than the number of observed seismicity rate, respectively. Dotted black line indicates the shut-in moment. \textbf{c.} Same as \textbf{b.} for HySei model. \textbf{d.} Same as \textbf{a.} for the Soultz-sous-For\^ets 2004 project. Note the different color scale. \textbf{e.} Same as \textbf{b.} for the Soultz-sous-For\^ets 2004 project. \textbf{f.} Same as \textbf{c.} for the Soultz-sous-For\^ets 2004 project.}	
\label{fig5}
\end{figure}

\begin{figure}[h!]
	\caption{Snapshots of magnitude frequency distribution of observed and forecast earthquakes (forecast is normalized so that total number of forecast earthquakes is equal to the number of observed events). Solid squares denote observed seismicity, colors refer to the same moments as on the previous two figures. Orange dots show SaSS forecast rate, blue dots indicate the HySei forecast rate. Corresponding transparent shaded areas indicate the $95\%$ confidence interval of the forecasts. Greenish shaded area indicates the overlapping of the two confidence intervals. \textbf{a.} Seismicity of the $3$-day forecast period after $1.25$ day of learning period. \textbf{b.} Same as \textbf{a.} after $3.25$ days of learning period. Inset shows a zoom of the black rectangle highlighting the magnitude bin of the largest event. \textbf{c.} Same as \textbf{b.} after $5.25$ days of learning period. \textbf{d.} Same as \textbf{a.} after $9.5$ days of learning period. \textbf{e.} Same as \textbf{a.} after $1.75$ days of learning period in Soultz-sous-For\^ets 2004. Note that last magnitude bin contains all magnitudes higher than $1.8$. \textbf{f.} Same as \textbf{e.} after $3.50$ days of learning period. \textbf{g.} Same as \textbf{e.} after $5.00$ days of learning period. \textbf{h.} Same as \textbf{e.} after $6.50$ days of learning period.}
	\label{fig6}
\end{figure}

\begin{figure}[h!]
	\caption{Space-test. Colorbar indicates joint log-likelihood values; yellow indicates better forecasts than red. Crosses represent moments when the model does not pass the Space-test. Gray squares denote moments when no earthquake occurred. Gray dotted line marks the shut-in moment. \textbf{a.} Space-test of SaSS for Basel 2006. \textbf{b.} Space-test of HySei for Basel 2006. \textbf{c.} Same as \textbf{a.} for Soultz-sous-For\^ets 2004. \textbf{d.} Same as \textbf{b.} for Soultz-sous-For\^ets 2004.}
\label{fig65}
\end{figure}

\begin{figure}[h!]
	\caption{Comparisons of model components based on log-likelihood. Green indicates moments when SaSS is superior to HySei, red shows moments when HySei performs better than SaSS. White denotes moments when both models perform similarly. Dotted black line indicates the shut-in moment. Note that scales are different for each component. \textbf{a.} Comparison of number components for Basel 2006. \textbf{b.} Comparison of magnitude components for Basel 2006. \textbf{c.} Comparison of spatial components for Basel 2006. \textbf{d.} Same as \textbf{a.} for Soultz-sous-For\^ets 2004. \textbf{e.} Same as \textbf{b.} for Soultz-sous-For\^ets 2004. \textbf{f.} Same as \textbf{c.} for Soultz-sous-For\^ets 2004.}
\label{fig7}
\end{figure}

\begin{figure}[h!]
	\caption{Cumulative joint LL values of the entire model summed for the indicated forecast periods and divided by the total number of observed events in the given forecast period. SaSS is denoted by solid orange dots, HySei is shown by blue circles. Dotted black lines show the shut-in moment. All values are plotted at the end of the corresponding learning period. Regime $A$ indicates the period when SaSS and HySei perform similarly, regime $B$ indicates the period when SaSS performs better than HySei, regime $C$ indicates the period when HySei performs better than SaSS. \textbf{a.} Cumulative joint LL/Eqk for $6$-hour forecast periods of the Basel 2006 experiment. \textbf{b.} Same as \textbf{a.} for $24$-hour forecast periods. \textbf{c.} Same as \textbf{a.} for $48$-hour forecast periods. \textbf{d.} Same as \textbf{a.} for $72$-hour forecast periods. \textbf{e.} Same as \textbf{a.} for Soultz-sous-For\^ets 2004.  \textbf{f.} Same as \textbf{b.} for Soultz-sous-For\^ets 2004.  \textbf{g.} Same as \textbf{c.} for Soultz-sous-For\^ets 2004.  \textbf{h.} Same as \textbf{d.} for Soultz-sous-For\^ets 2004.}
	\label{fig8}
\end{figure}

\begin{figure}[h!]
	\caption{Comparison of different methods to evaluate average information gain (HySei is the reference model). Solid black line shows the empirical probability distribution of the information gain values, dotted gray line indicates the normal distribution, which expected value and standard deviation is calculated from the information gain population. Dashed red, dashed green, solid orange, and solid brown lines correspond to classical mean, robust mean, bootstrap mean, and bootstrap median, respectively. Insets show the estimated average values with their uncertainties. Reddish background denotes the area, where SaSS is better, yellowish background denotes the area, where HySei is better. The number of investigated earthquakes are shown in the top left corner of the graph. \textbf{a.} Information gain in the case of the Basel 2006 experiment. \textbf{b.} Same as \textbf{a.} for Soultz-sous-For\^ets 2004.}
	\label{fig9}
\end{figure}

\begin{figure}[h!]
	\setcounter{figure}{0}
	\includegraphics[scale = 0.45]{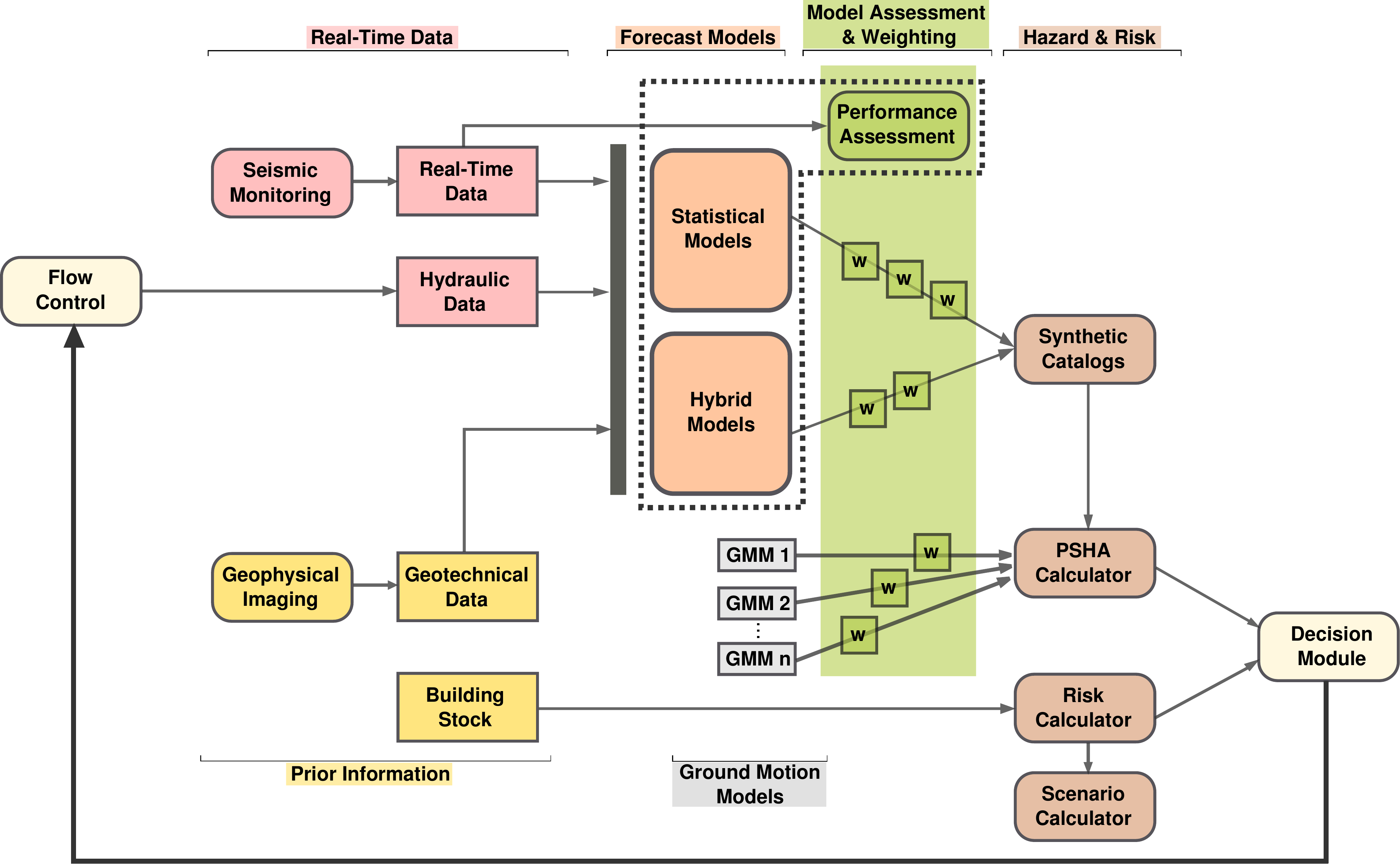}
	\caption{}
	\label{fig0}
\end{figure}

\begin{figure}[h!]
	\setcounter{figure}{1}
	\includegraphics[scale = 0.6]{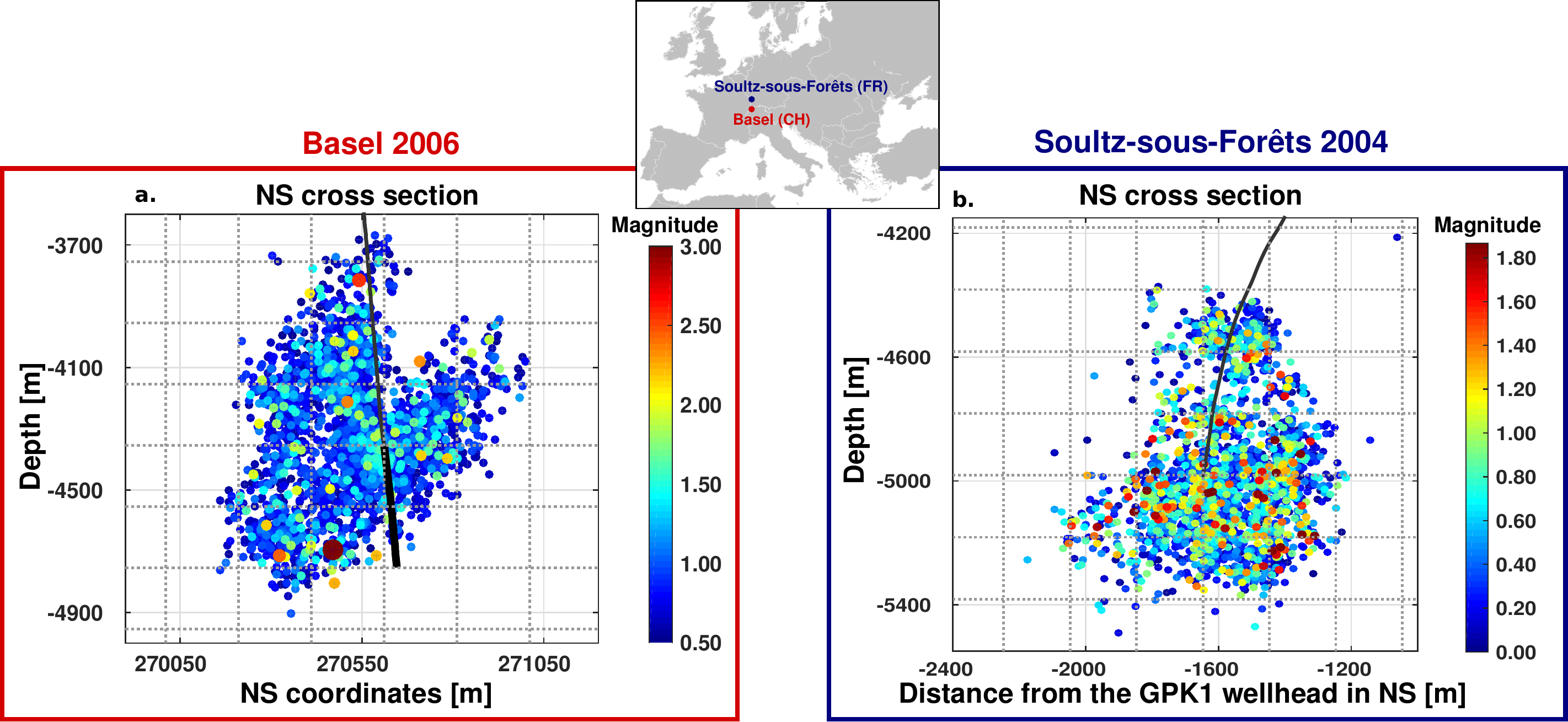}
	\caption{}
	\label{fig1}
\end{figure}

\begin{figure}[h!]
	\setcounter{figure}{2}
	\includegraphics[scale=0.65]{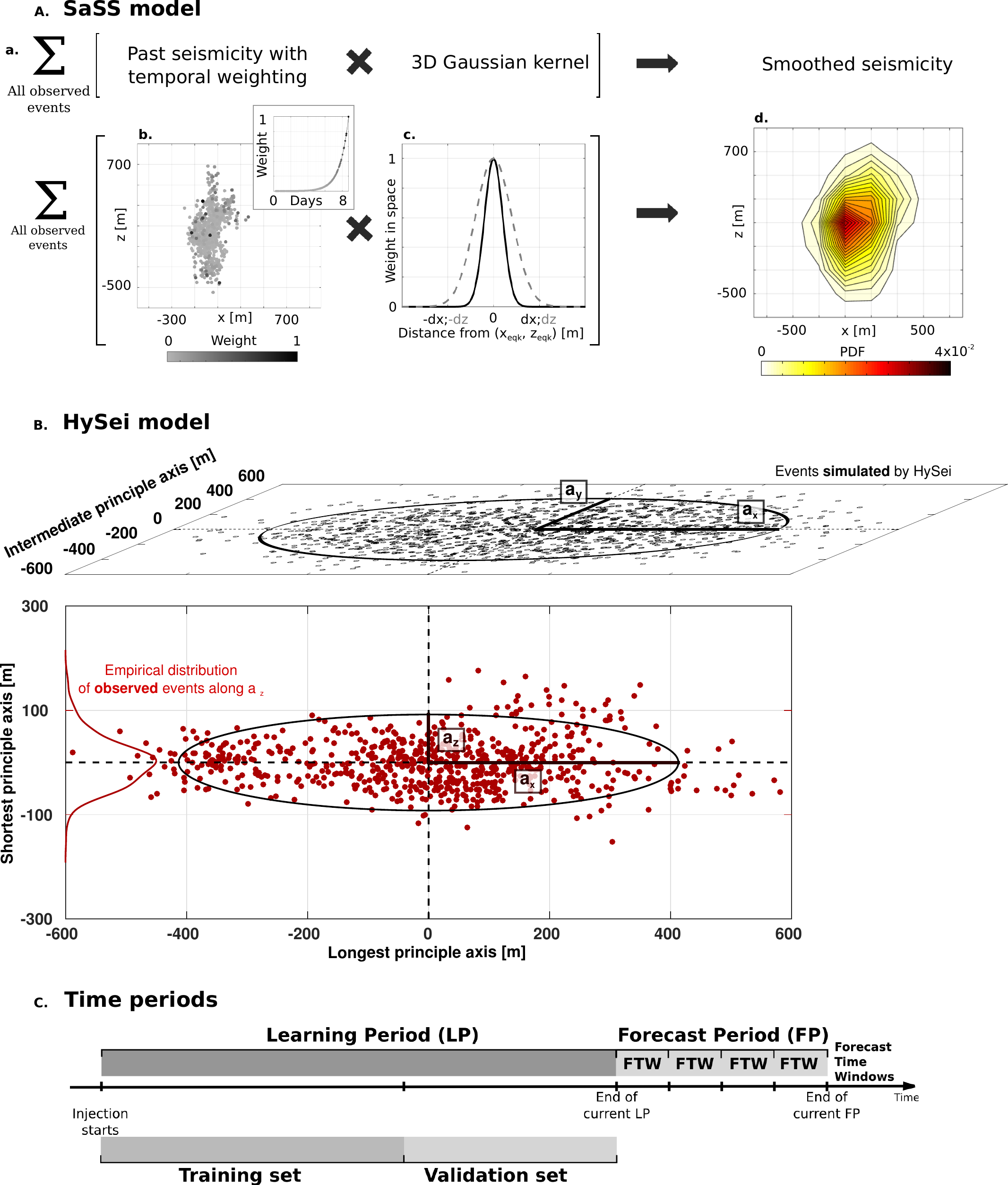}
	\caption{}
	\label{fig2}
\end{figure}

\begin{sidewaysfigure}
	\setcounter{figure}{3}
	\includegraphics[scale = 0.6]{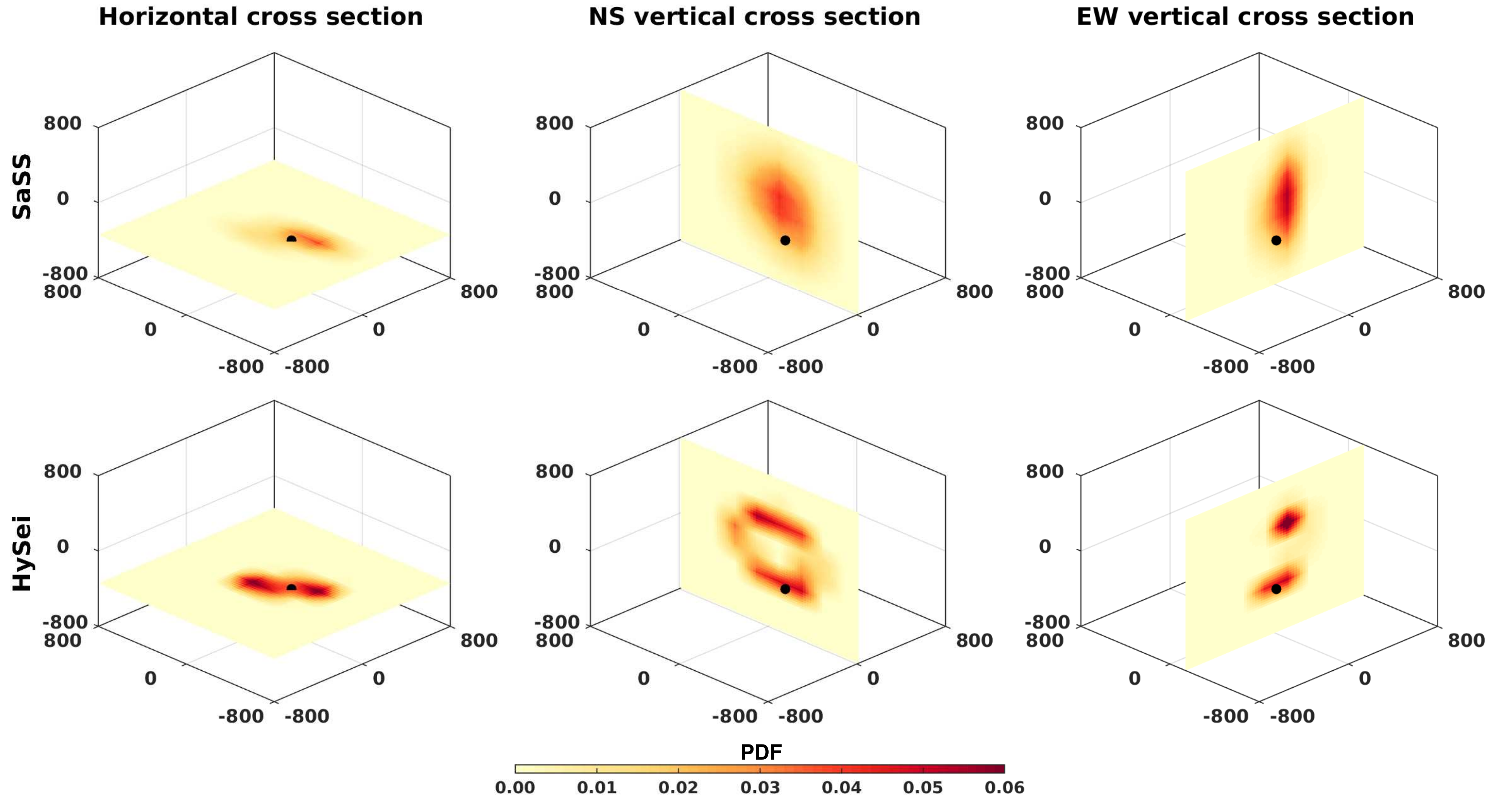}
	\caption{}
	\label{fig3}	
\end{sidewaysfigure}

\begin{sidewaysfigure}[h!]
	\setcounter{figure}{4}
	\includegraphics[scale = 0.55]{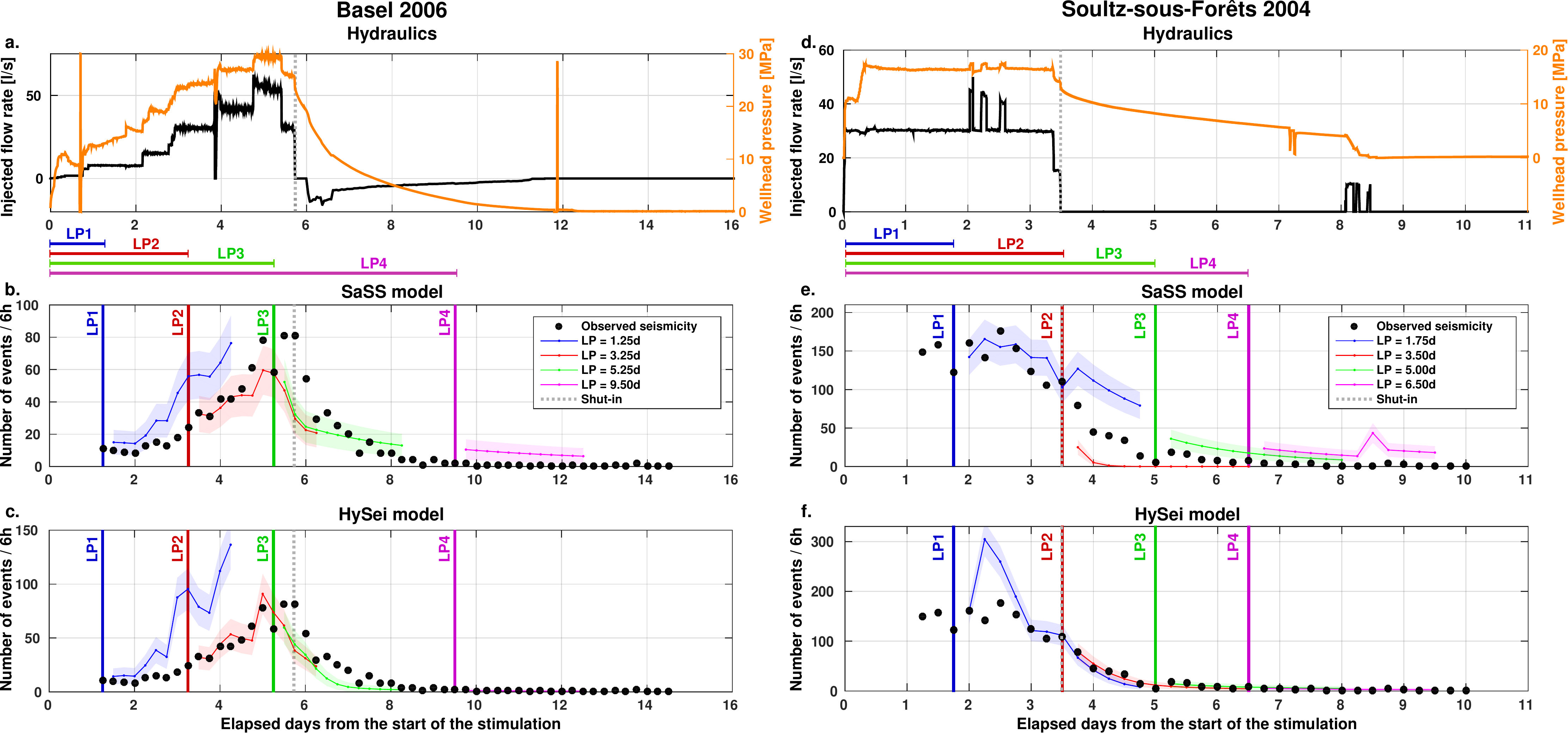}
	\caption{}
	\label{fig4}	
\end{sidewaysfigure}

\begin{sidewaysfigure}
	\setcounter{figure}{5}
	\includegraphics[scale = 0.7]{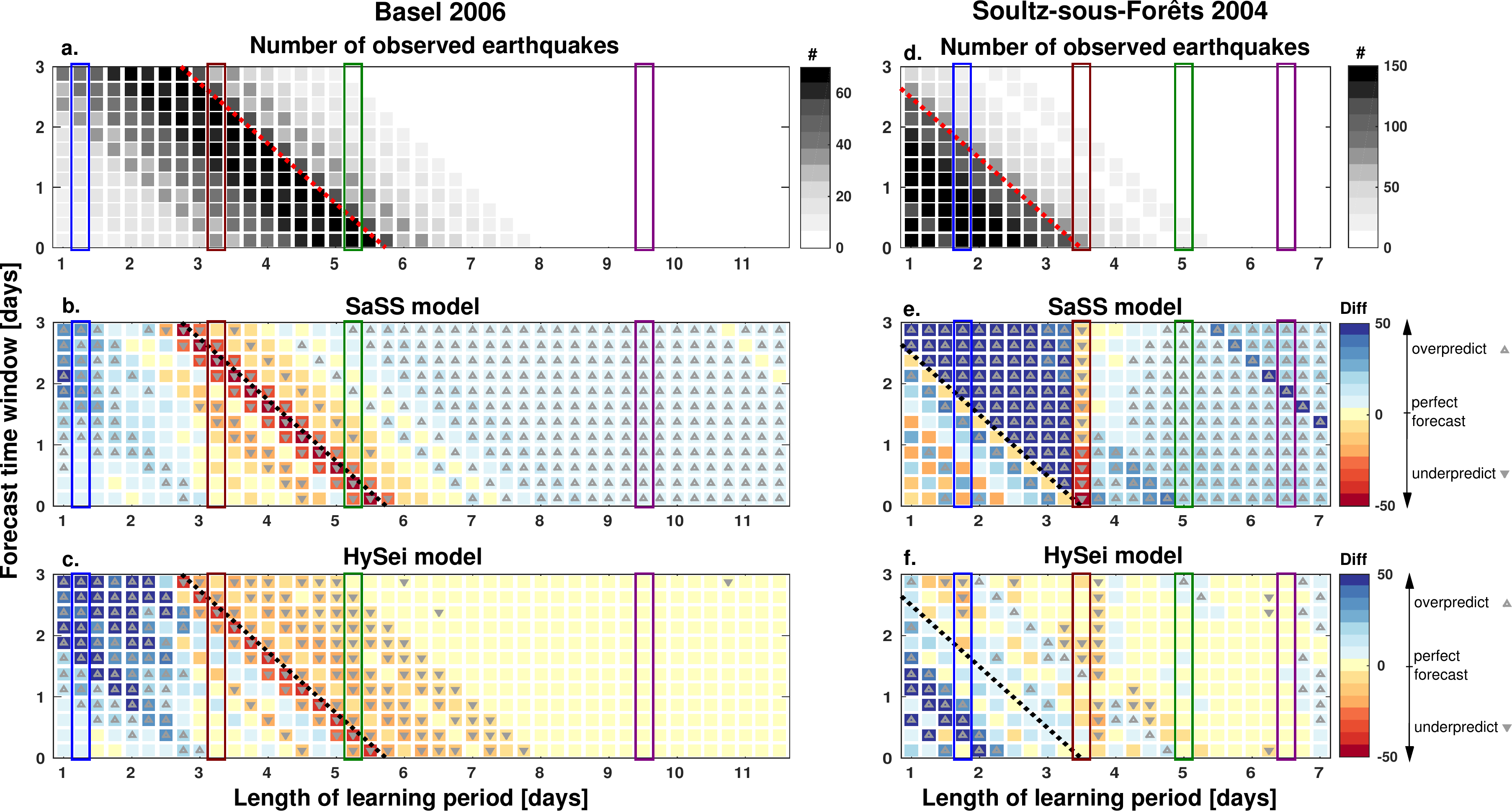}
	\caption{}
	\label{fig5}
\end{sidewaysfigure}

\begin{figure}[h!]
	\setcounter{figure}{6}
	\includegraphics[scale = 0.6]{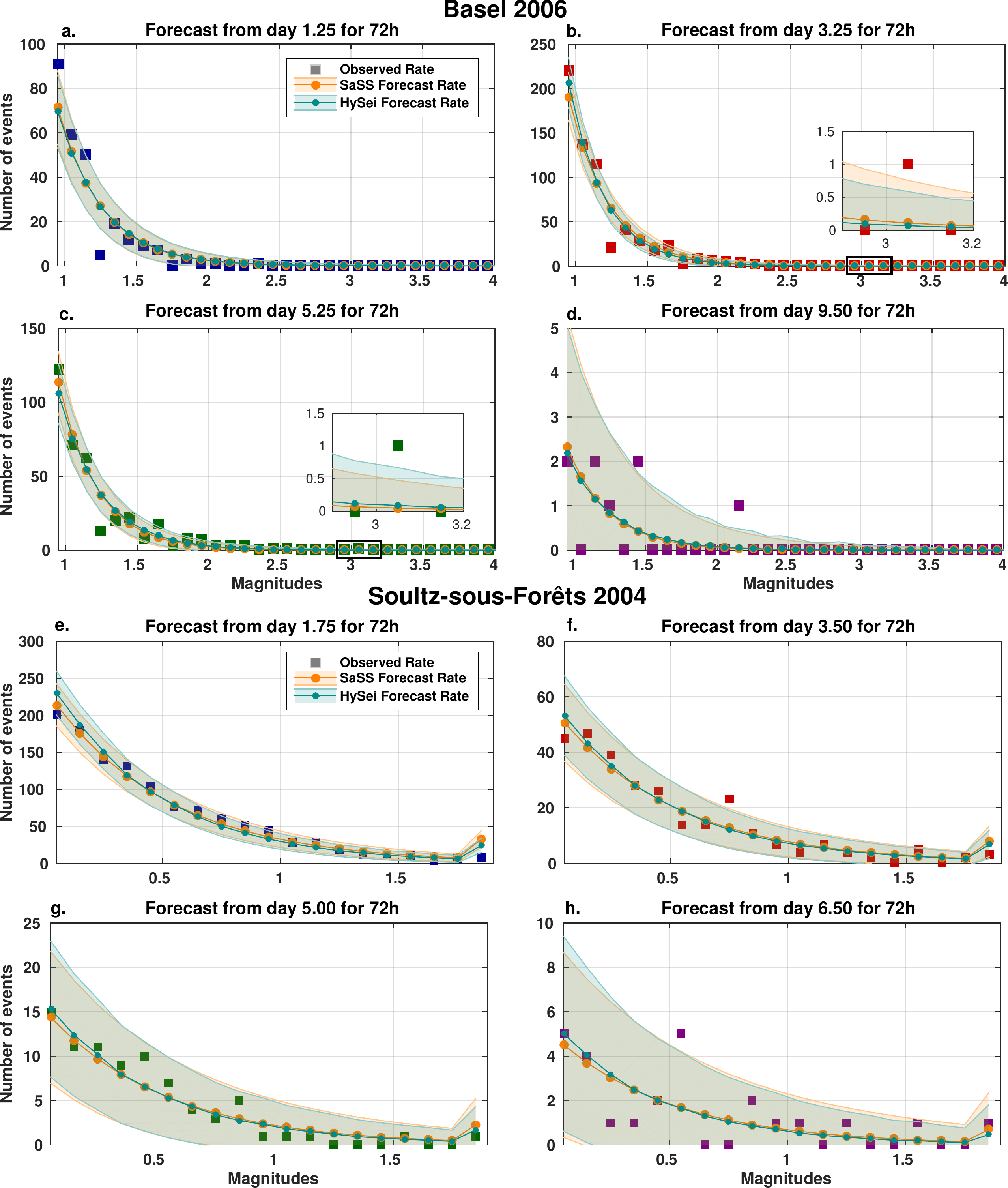}
	\caption{}
	\label{fig6}
\end{figure}

\clearpage
\begin{sidewaysfigure}
	\setcounter{figure}{7}
	\includegraphics[scale = 0.85]{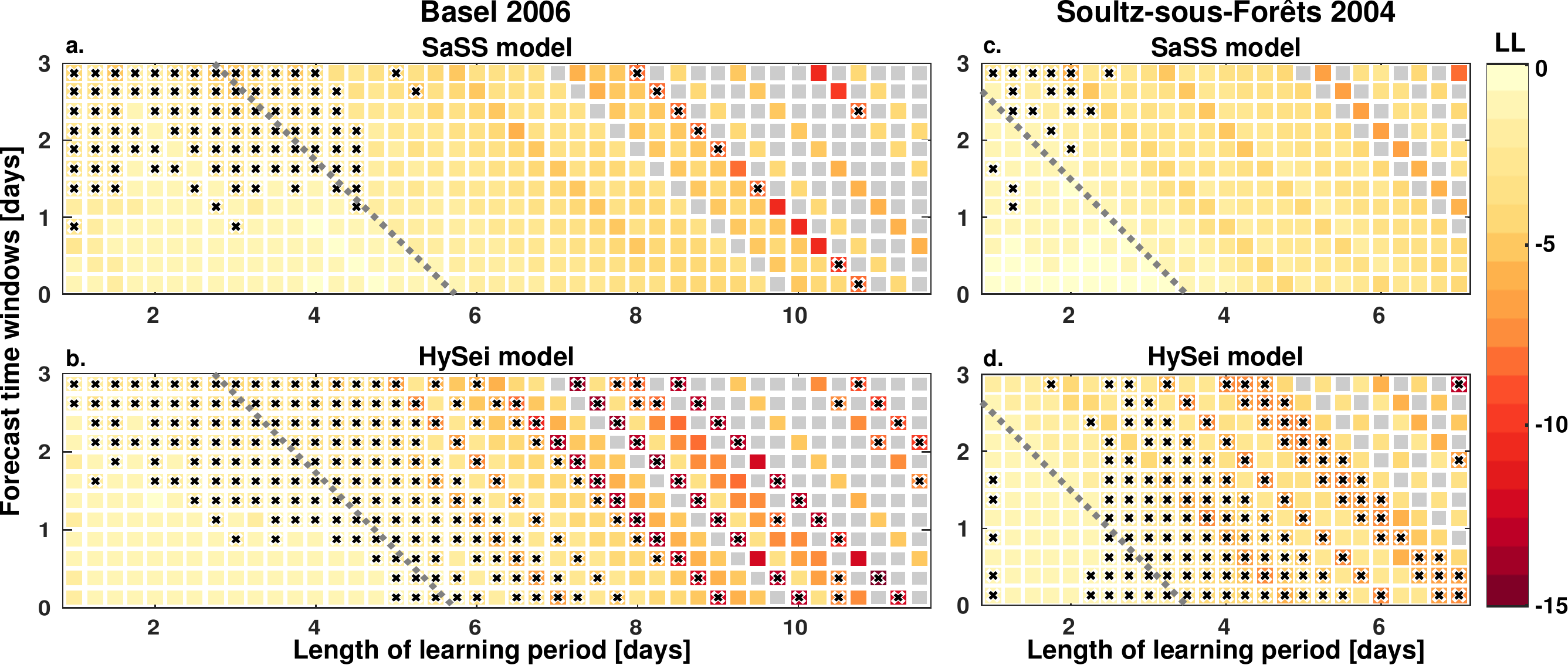}
	\caption{}
	\label{fig65}
\end{sidewaysfigure}

\clearpage
\begin{sidewaysfigure}
	\setcounter{figure}{8}
	\includegraphics[scale = 0.75]{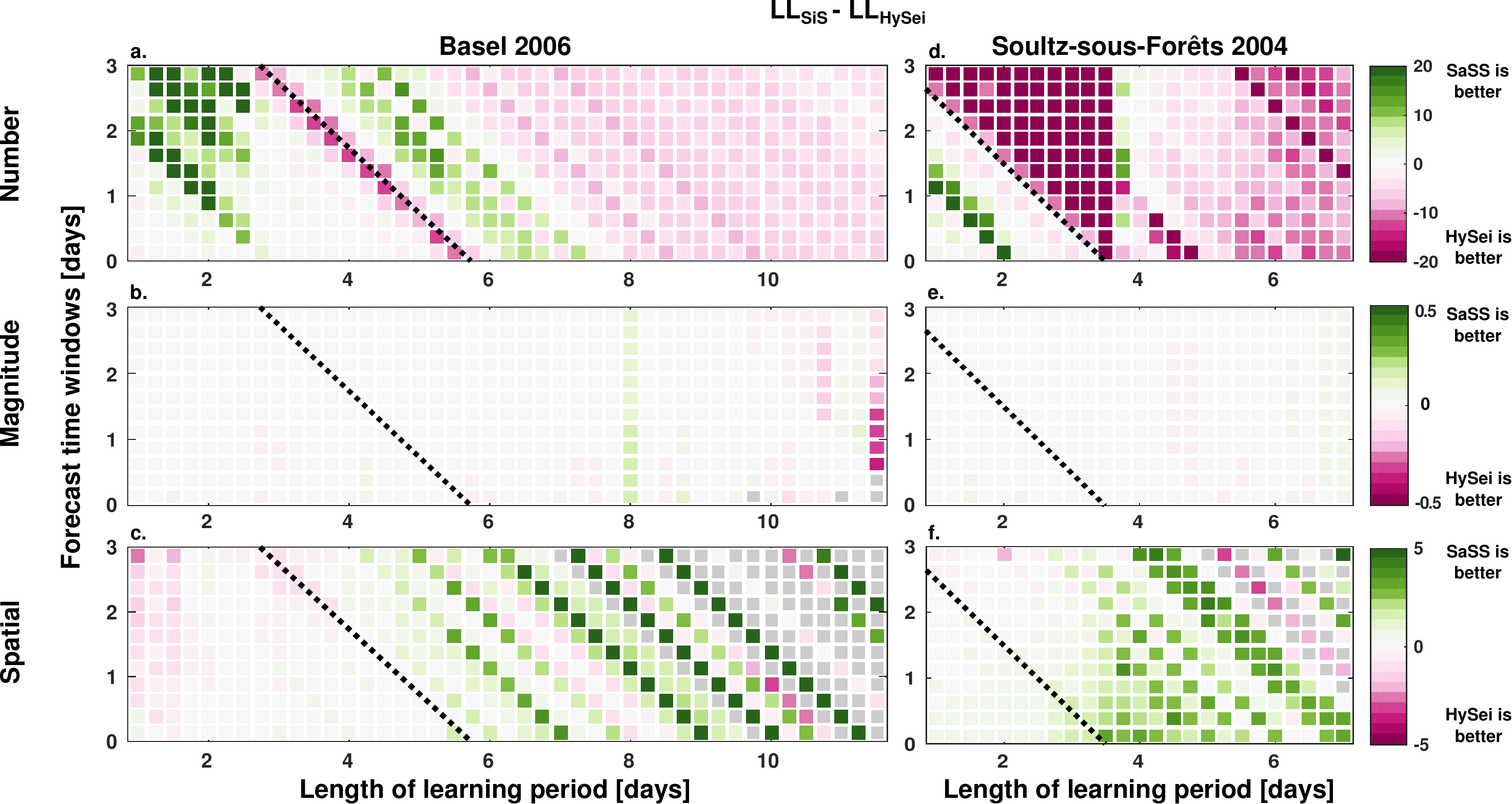}
	\caption{}
	\label{fig7}
\end{sidewaysfigure}

\begin{figure}[h!]
	\setcounter{figure}{9}
	\includegraphics[scale = 0.7]{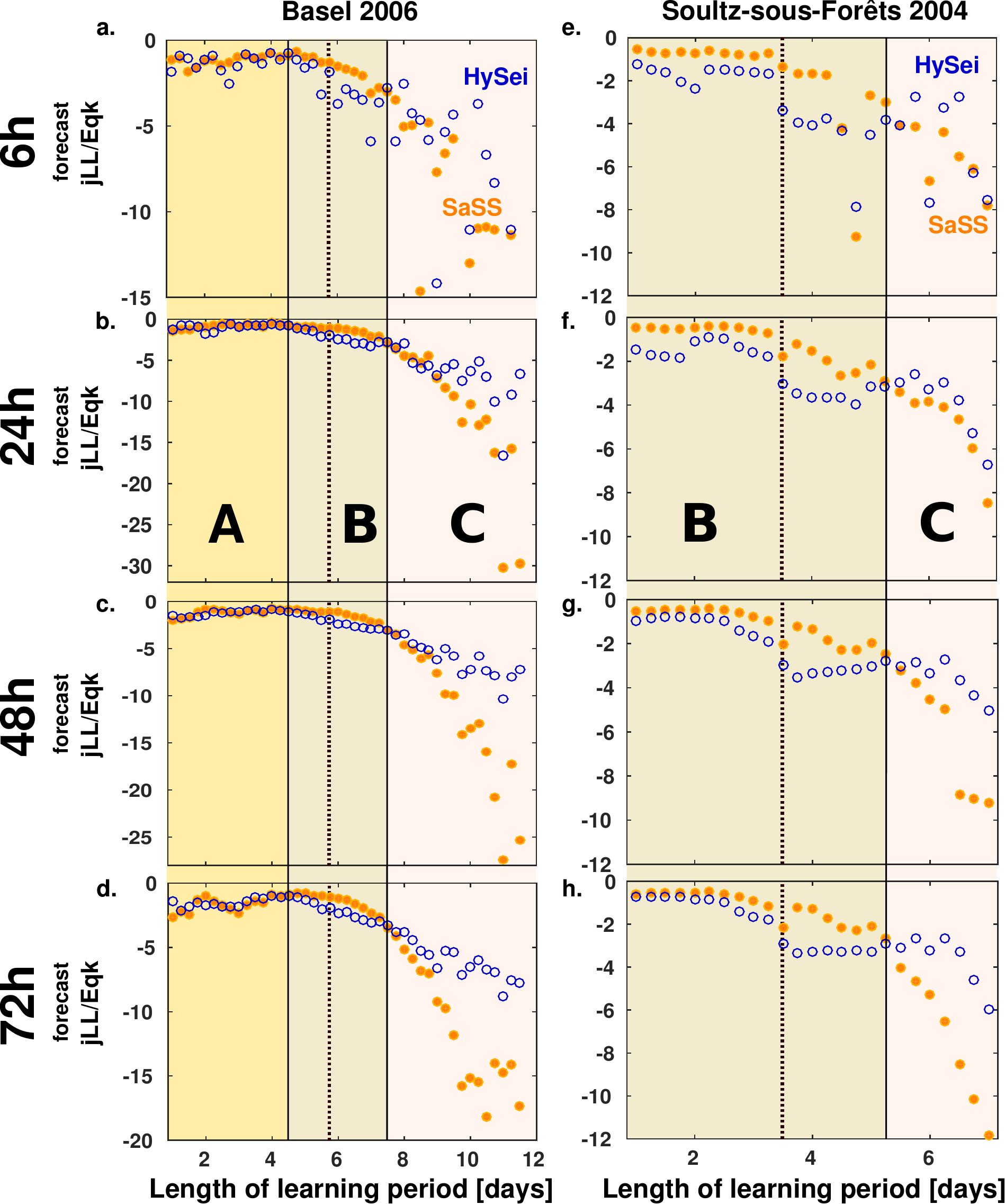}
	\caption{}
	\label{fig8}
\end{figure}

\begin{figure}[h!]
	\setcounter{figure}{10}
	\includegraphics[scale = 0.65]{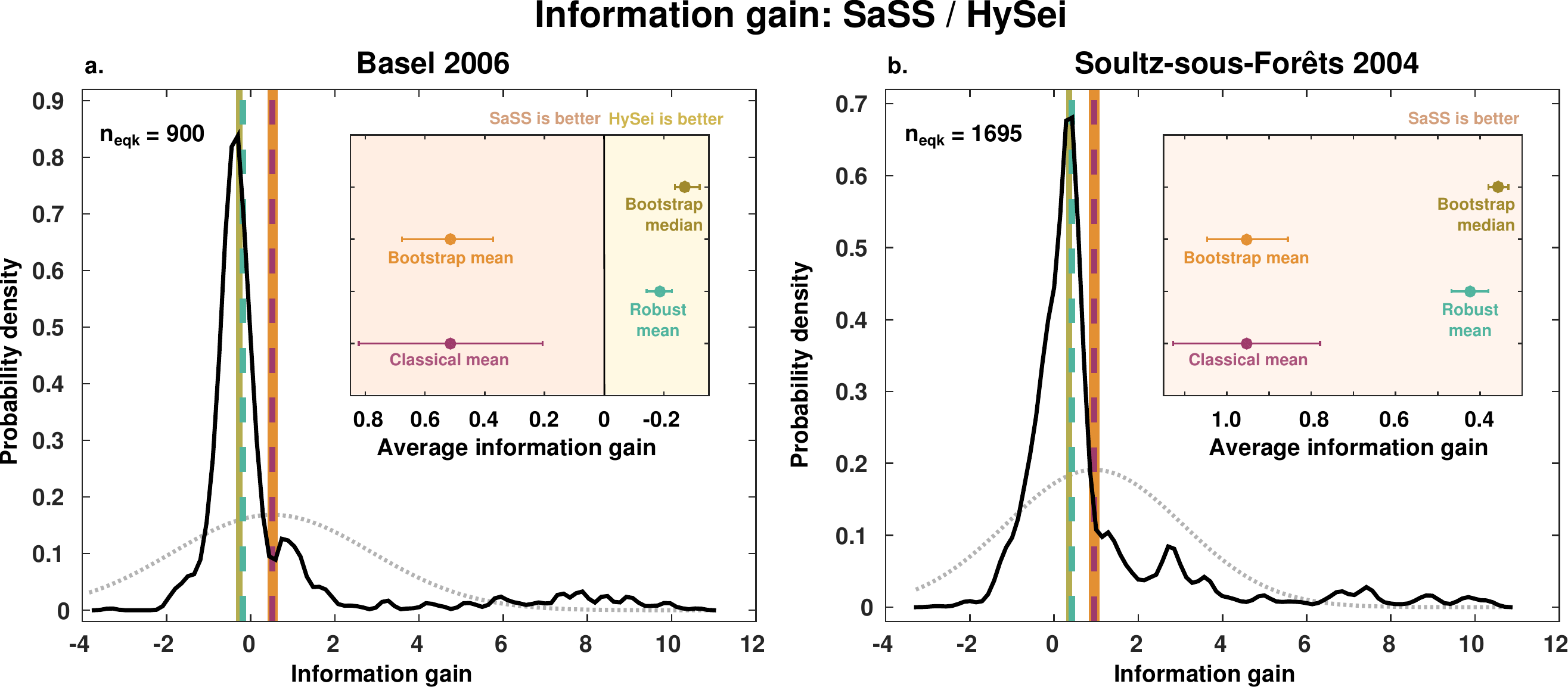}
	\caption{}
	\label{fig9}
\end{figure}

\end{document}